\documentclass[a4paper,conference]{IEEEtran}
\usepackage{multirow}
\usepackage{amsmath,graphicx}
\usepackage{url}            
\usepackage{booktabs}       
\usepackage{amsfonts}       
\usepackage{nicefrac}       
\usepackage{microtype}      
\usepackage{algorithm}
\usepackage{algpseudocode}  
\usepackage{caption}
\usepackage{subfigure}
\usepackage{enumitem}

\ifCLASSINFOpdf
\else
\fi
\begin{document}
%
\title{CAggNet: Crossing Aggregation Network for Medical Image Segmentation}

\author{
\IEEEauthorblockN{Xu Cao}
\IEEEauthorblockA{School of Data Science\\ 
Fudan University\\
Shanghai, China\\
Email: caox16@fudan.edu.cn}
\and %
\IEEEauthorblockN{Yanghao Lin}
\IEEEauthorblockA{School of Data Science\\
Fudan University\\
Shanghai, China\\
Email: linyh16@fudan.edu.cn}
}


%


\maketitle

\begin{abstract}
In this paper, we present Crossing Aggregation Network (CAggNet), a novel densely connected semantic segmentation approach for medical image analysis. The crossing aggregation network improves the idea from deep layer aggregation and makes significant innovations in semantic and spatial information fusion. In CAggNet, the simple skip connection structure of general U-Net is replaced by aggregations of multi-level down-sampling and up-sampling layers, which is a new form of nested skip connection. This aggregation architecture enables the network to fuse both coarse and fine features interactively in semantic segmentation. It also introduces weighted aggregation module to up-sample multi-scale output at the end of the network. We have evaluated and compared our CAggNet with several advanced U-Net based methods in two public medical image datasets, including the 2018 Data Science Bowl nuclei detection dataset and the 2015 MICCAI gland segmentation competition dataset. Experimental results indicate that CAggNet improves medical object recognition and achieves a more accurate and efficient segmentation compared to existing improved U-Net and UNet++ structure.
\end{abstract}


%
\IEEEpeerreviewmaketitle
\section{Introduction}
Semantic segmentation is one of the core jobs of image processing and analysis. During the segmentation process, original images or videos are separated into different regions based on plenty of semantic features and recover to pixel-wised probability maps at the end of the model. These features can be extracted from pixel intensity value, crossing pattern, geometric shape, etc\cite{garcia2017review}. In general, image semantic segmentation by deep learning can be defined as detecting pixel-wise region categories by CNN-based network's parameter training.

\begin{figure}[h]
\centering
\subfigure[] 
{
	\centering          
	\includegraphics[width=0.35\linewidth]{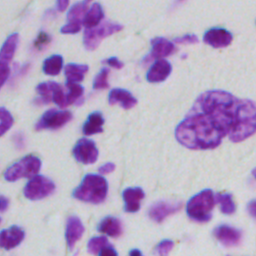}   
}
\subfigure[]
{
	\centering     
	\includegraphics[width=0.35\linewidth]{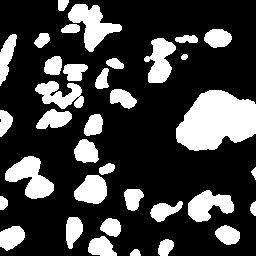}  
}
\caption{Example of medical image segmentation job: nuclei detection of cells}
\label{fig:dataset_example}
\end{figure}

Among plenty of semantic segmentation jobs, medical image segmentation is one of the most challenging tasks due to its lack of samples and labels. In clinical research and application, medical image segmentation is a significant procedure for clinical evaluation and diagnosis. It includes all kinds of segmentation tasks from the 2D cell level to the 3D organ and system level of human bodies. Fig.\ref{fig:dataset_example} shows an example of detecting cell nuclei in microscopy images. The shape and area of unusual objects in medical images such as CT images, nuclear magnetic images, and microscopy images can offer clinicians crucial insight into patients' severity and plan for treatment. It has become a consensus of clinical researchers using computers to assist in the diagnosis of medical images nowadays.

When fighting with COVID-19 around the world, the gold standard for diagnosis is real-time polymerise chain reaction (RT-PCR) assay of the sputum. However, using RT-PCR is time-consuming especially when repeated verification is required. In fact, radio-logical reports derived from CT image segmentation can also help clinician to analyze COVID-19 lesions in the lungs. Some new studies showed that medical image segmentation methods based on deep learning can quickly detect these COVID-19 lesion regions and provide enough information for clinical to evaluate and diagnose\cite{fan2020inf, shan2020lung}. They have broad application prospects to become auxiliary strategy of COVID-19 diagnosis.

Recently, Deep Convolutional Neutral Network (DCNN) gradually replaced traditional computer vision methods to solve medical image segmentation tasks. In the medical image object detection area, Faster R-CNN\cite{ren2015faster} is gradually becoming the backbone of various new models. Similarly, in the medical image segmentation area, the ideas of U-Net\cite{ronneberger2015u} and ResNets\cite{he2016deep} have also been absorbed by many new medical image semantic segmentation networks. Obviously, the exceptional performance and better robustness of deep learning approaches have yielded great influence in medical image and other interdisciplinary research\cite{garcia2017review}. However, due to the low resolution and blurred boundary of medical images, it is still a challenging task to design new models that can more effectively capture more fine-grained details.

In order to implement the request for more accurate segmentation in medical images, we propose Crossing Aggregation Network (CAggNet) that incorporates nested skip connection from residential blocks from ResNets, UNet++\cite{zhou2018unet++} and DLA\cite{yu2018deep}. We introduce two sub-structures: crossing aggregation module (CAM) and weighted aggregation module (WAM). CAM focuses on merging features from different layers on different scale levels while WAM focuses on aggregating multi-scale output information. CAggNet can fully embeds dense skip connections into U-Net's encoder-decoder structure. 

Our main contributions are summarized as follows:

\begin{itemize}
\item Propose a crossing aggregation module to densely and iteratively integrate information in the U-Net structure.
\item Redesign a graph-based deep layer aggregation method by combining crossing aggregation and weighted aggregation.
\item Introduce focal loss in deep layer aggregation framework and prove its robustness in our model.
\end{itemize}

The next section reviews several classic medical image segmentation research-related work. Then, section 3 provides a detailed description of the Crossing Aggregation Network (CAggNet). Section 4 presents and compares the experimental results among our CAggNet and some baseline segmentation methods such as U-Net. Our conclusions and summaries are provided in the final section.

\section{Related Work}
After the discovery of fully convolutional networks (FCN)\cite{long2015fully}, U-Net\cite{ronneberger2015u} and residual networks (ResNets)\cite{he2016deep}, a large number of image segmentation framework in Kaggle medical image segmentation competition have adopted their skip connection and residual block idea. FCN originated from the improvement of classification network: VGG-Net\cite{garcia2017review}. In FCN, the last fully connected layers of the VGG-Net is replaced by convolutional layers to output a probability mask of the input image. The U-Net architecture further applies multi-layer skip-connection to combine low-level feature maps from the encoder sub-network with higher-level feature maps from the decoder sub-network, which enables a more precise pixel-level localization. The skip connection in U-Net can restore the information loss caused by the down-sampling process, which produces similar effects to the residual block in ResNets.

Recently, a series of U-Net improvement architectures came into researchers' sight. Ternaus-Net\cite{iglovikov1801u} applied the idea from transfer learning to replace the encoder part of U-Net by a pre-trained VGG-11 network. W-Net\cite{xia2017w} achieved a two-stage segmentation by concatenating together
two FCN/U-Net into an encoder-decoder framework. Res U-net\cite{xiao2018weighted} and Dense U-net\cite{guan2019fully} replaced each U-Net's encoder block and decoder block with residual connections and dense connections while attention U-Net\cite{oktay2018attention} added self-attention block to optimize the skip connection pathway of U-Net. Despite the substructure of these models are different, they all share a similar linear encoder-decoder U-Net backbone.

Although U-Net has skip-connection pathways to combine features, its direct fusion of semantic information can not achieve good results on issues or dataset with blurred and sophisticated shape\cite{zhang2018exfuse}. One improvement idea here is to apply a nested structure to reduce the semantic gap between the feature maps of the encoder and decoder sub-networks, including deep layer aggregation (DLA)\cite{yu2018deep}, UNet++\cite{zhou2018unet++} and UNet3+\cite{huang2020unet}. Both deep layer aggregation and UNet++ absorb the idea of nested skip connection from densely connected networks (DenseNets)\cite{huang2017densely}. DenseNets are a series of improved architectures of ResNets that concatenate all the layers in different stages by propagating features and losses through skip connections\cite{huang2017densely}. Like ResNets, DenseNets can also achieve more accuracy by deepening the network to achieve more accuracy as well as obtain better parameters and better memory efficiency. Deep layer aggregation structure iteratively and hierarchically merge the feature by a group of convolutional layers in each depth of the encoder-decoder structure. Similarly, UNet++/UNet3+ connects the encoder and decoder sub-network through a series of nested dense convolutional blocks. The core hypothesis of all these architectures is that the nested encoder-decoder model can capture some more fine-grained details of the foreground objects from the shallower network level to the high-resolution feature maps. In contrast to general U-Net structure, DLA and UNet++ aggregate richer information for recognition and localization on skip pathways. These models' segmentation results show better fusions of semantic information and more precise feature extraction.

To strengthen the skip connection and improve the efficiency of U-Net's encoder-decode structure, we present CAggNet, a new segmentation architecture based on a densely crossing connection method to replace traditional skip connections. This model can address the need for more accurate segmentation in complex medical images.

\section{Proposed Network}

\begin{figure*}[h]
  \centering
  \includegraphics[width=1.0\linewidth]{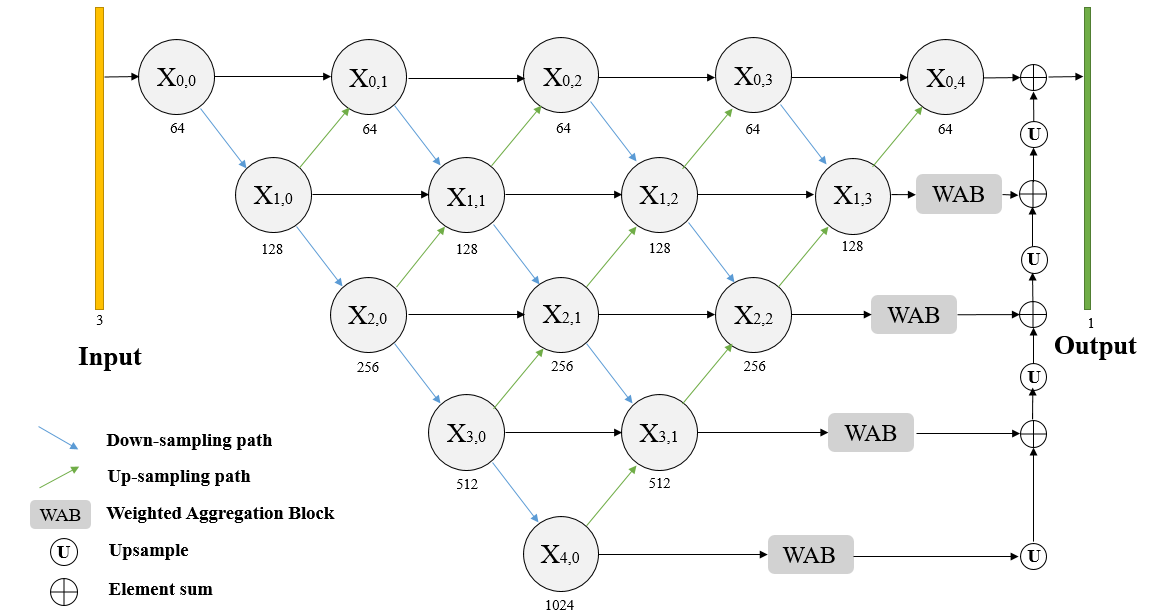} 
  \caption{The proposed CAggNet architecture.}
  \label{fig:ca_architecture}
\end{figure*}

Fig.\ref{fig:ca_architecture} illustrates the overall framework of our crossing aggregation network, which is primarily based on an encoder-decoder architecture, with multiple convolutional layers in each resolution level's skip connection pathways. The network can be divided into two sub-structure: crossing the aggregation module and weighted aggregation module. In crossing aggregation module, the parameters are transferred among convolutional layers through up-sampling, down-sampling and concatenate operation. Then, the weighted aggregation module merges the outputs of the cross-aggregation module from different resolution level in order to better recover the final probability predicted map. The model introduction comprises three parts: 1) the basis of our research: deep aggregation, 2) the structure of crossing aggregation module(CAM), and 3) weighted aggregation module(WAM).

\subsection{Deep Aggregation}

Since the skip connections in U-Net are linear and only merge resolution maps in the same layer, some significant semantic and spacial information does not fuse well enough. This makes U-Net not suitable for training in segmentation dataset containing too much semantic information. To handle this problem, Yu et al.\cite{yu2018deep} introduce a method called deep layer aggregation (DLA), which shares a similar idea with UNet++. The basic idea of DLA is to combine different layers throughout a network. It contains two sorts of aggregation: Iterative Deep Aggregation (IDA) and Hierarchical Deep Aggregation (HDA)\cite{yu2018deep}. IDA is an iterative structure that progressively aggregates and fuses resolution maps/feature maps by extending additional layers between the original skip connection path. Its role is to refine and enhance semantic information fusion in the skip connection. HDA embodies the idea of Semantic Embedding. It produces a tree structure which makes shallow and deep layers become closer spatially\cite{kang2019nuclei}.

\begin{figure}[h]
\centering
{
	\centering          
	\includegraphics[width=1.0\linewidth]{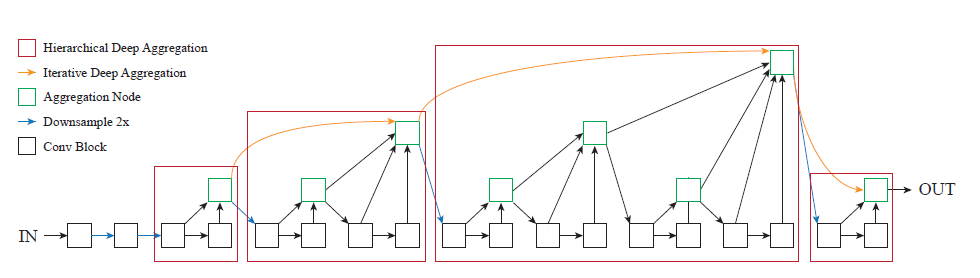}   
}
\caption{Deep layer aggregation consists of IDA (orange arrow) and HDA (red box)\cite{yu2018deep}. This tree-shaped encoder-decoder structure has been proven to be robust in nuclei segmentation task\cite{kang2019nuclei}}
\label{fig:deep_aggregation}
\end{figure}

Fig.\ref{fig:deep_aggregation} shows the architecture of DLA combining by IDA and HDA. IDA connections join adjacent layers to gradually deepen and spatially refine features' representation. HDA connections use a tree structure to involve more semantic information from high-level features fusing into low-level features.

\subsection{Crossing Aggregation Module}

\begin{figure}[h]
\centering
{
	\centering          
	\includegraphics[width=1.0\linewidth]{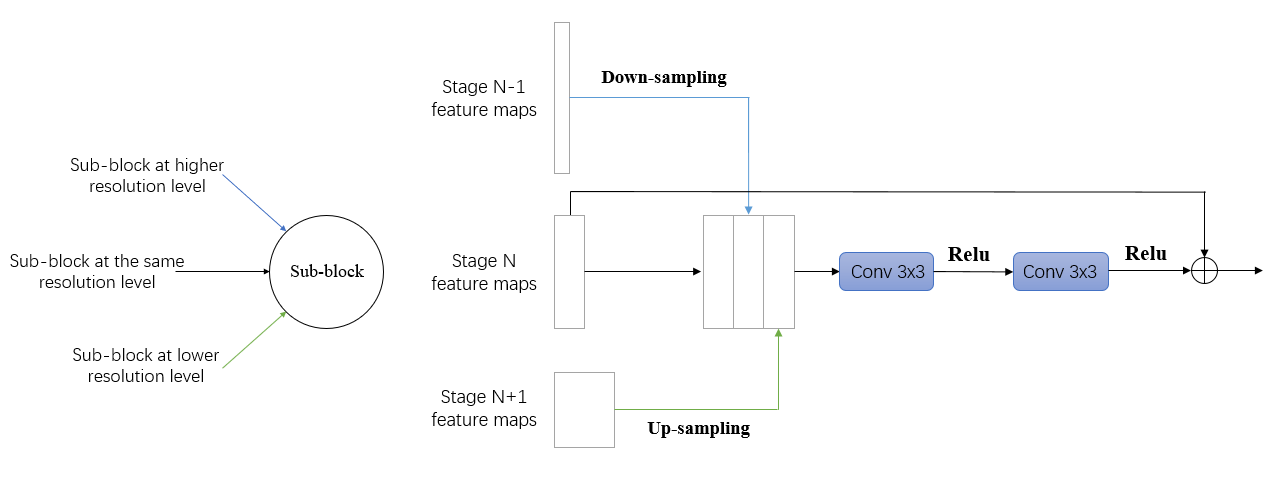}   
}
\caption{Crossing Aggregation Module.}
\label{fig:ca_module}
\end{figure}

Although deep layer aggregation performs well in many semantic segmentation tasks, its tree shape iterative fusion method makes the advantage of the original U-Net direct skip connection weakened. In UNet++, Zhou et al.\cite{zhou2018unet++} apply dense connections to keep the skip connection path still available after adding aggregation layers. However, the application of dense connections also means that the training process of the UNet++ is difficult. Although, Zhou et al. apply many strategies like pruning to improve model training, these defects still make UNet++ weaker than the original U-Net in terms of model versatility and stability.

In CAggNet, we redesign a novel aggregation module: Crossing Aggregation Module (CAM), which can be considered as a graph based full scale layer aggregation approach. The input of CAM contains three different scale level feature maps: stage N-1 feature map(higher level resolution), stage N feature map, and stage N+1 feature map(lower level resolution). As an example, Fig.\ref{fig:ca_module} shows how to pass parameter in one CAM block. CAM first concatenates the stage N feature map, down-sampling of stage N-1 feature map and up-sampling of stage N+1 feature map. Then, it passes through two 3$\times$3 convolutional layers connected with the Relu function. Finally, the output feature map of the last Relu function is summed up with the input stage N-1 feature map. The practical effect of CAM is to use the idea of residual blocks to fuse different pyramid levels of information while the connection of CAMs can also iteratively deepens adjacent resolution level of the network. Using CAM, the model can connect all resolution and scale level in a graph structure, which may achieve a better multi-scale feature combination than UNet++. Besides, similar to the effect of ResNets, this aggregation method can avoid the network from losing parameters due to the increase in depth, and it can also keep stable training process. The crossing aggregation module is defined by:

\begin{equation}
\begin{aligned}
Z=Concat(X_{i,j-1},DS(X_{i-1,j}),US(X_{i+1,j}))
\end{aligned}
\end{equation}

\begin{equation}
\begin{aligned}
X_{i,j}=X_{i,j-1}+\sigma_2(W_2\sigma_1(W_{1}Z+b_1)+b_2)
\end{aligned}
\end{equation}

where DS and US represent down-sampling and up-sampling operation respectively. $\sigma$ is the Relu activation, and $W$ and $b$ are the weights in the convolutional layer. CAM also apply the batch normalization operation to reduce generalization error and accelerate training.

In brief, CAggNet iteratively connects all CAM block, as Fig.\ref{fig:ca_architecture} shows. The aggregation of CAM can achieve a better skip connection merging process through a graph based structure. Further, we find that the composition of CAggNet can be split into multiple sub-encoder-decoder structures, such as sub-U-Net, sub-W-Net. It means that the connection between CAMs can also be considered as a tandem sub-encoder-decoder structure. This iterative network framework may has high research value for multi-stage semantic segmentation in the future.

\subsection{Weighted Aggregation Module}

\begin{figure}[h]
\centering
{
	\centering          
	\includegraphics[width=1.0\linewidth]{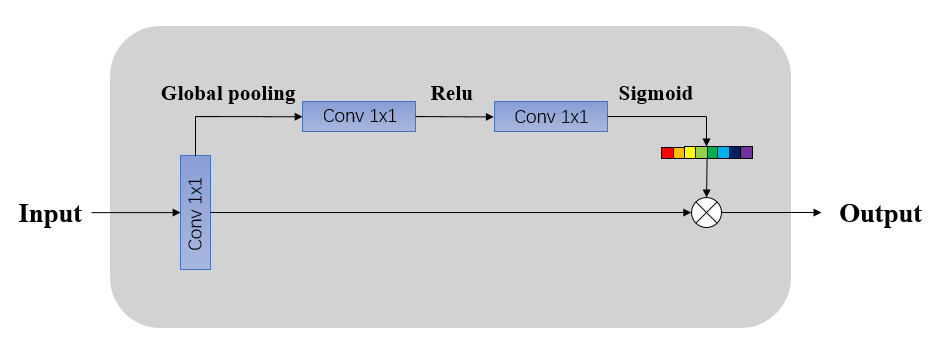}   
}
\caption{Weighted Aggregation Block(WAB).}
\label{fig:wag_module}
\end{figure}

Weighted Aggregation Module was first proposed by Zhang et al.\cite{zhang2019net} in ET-Net. It emphasizes aggregating valuable features and multi-scale information to improve segmentation performance. As shown in Fig.\ref{fig:wag_module}, the weighted aggregation block consists of a global pooling operation and two activation functions, which is equivalent to the idea of channel attention module\cite{zhang2019net}. In this block, the input feature maps first generate a one dimension attention vector by global average pooling. Then, the vector generates the channel attention weights by passing two 1$\times$1 convolutional layers with Relu and sigmoid activation functions. Finally, the output of the WAB is an element-wise product of input feature maps and channel attention vectors.

In CAggNet, WAB calculates channel attention vectors of the last layer of crossing aggregation in different scales and optimizes the output of the feature map through channel attention. At the same time, the output of each layer of WAB is combined through up-sampling and concatenate operations from bottom to up. This multi-level channel attention fusion method is conducive to strengthen the detection and recovery of object details in semantic segmentation prediction.

\subsection{Focal Loss}
Most of medical image segmentation problems can be defined as foreground-background class imbalanced problems. This is because there is often a lot of noise and background interference in medical images. Sometimes, the foreground objects that we need to detect are only distributed in a small region of input images. Cross entropy loss is not conducive to the training process of these imbalanced class situation because it will fail early in the network training (falling into a local minimum) or lead to degenerate models with poor performance. In order to handle these problems, Lin et al.\cite{lin2017focal} introduced focal loss, a method that is applied to train on sparse sets. The focal loss is defined as:

\begin{equation}
\begin{aligned}
FocalLoss(P_{t})=-\alpha_{t}(1-P_{t})^{\gamma}log(P_{t})
\end{aligned}
\end{equation}

where $P_{t}$ is predicted probability. $\alpha$ and $\gamma$ are two hyperparameters respectively denote for balanced variant of foreground and background classes and focusing parameter to rescale the loss. When $\gamma=0$, the focal loss is equivalent to Weighted BCE loss. If $\gamma$ increase, the effect of the modulating factor will also increase. Lin et al.\cite{lin2017focal} also found that focal loss work well in their experiments if $\gamma=2$ and $\alpha=0.5$.

\section{Experiment}
The models are implemented in Pytorch. Adam optimizer\cite{kingma2014adam} is used to optimize models and the learning rate is set to 1e-3. Batch size is set to 5 for the CELL dataset and 2 for the GLAND dataset. Early-stopping on the validation set is applied with a patience of 32 epochs. The models generally converge within 100 epochs. The IOU and F1-score are computed on the validation set. All models are trained on two NVIDIA GTX 1080Ti GPUs.

\subsection{Datasets and Evaluation Metrics}
We evaluated our method on two public datasets. The first dataset (referred to as CELL) is 2018 Data Science Bowl aiming to detect the nuclei in divergent microscopy cell images. In this dataset, there are 670 cell images in total, 509 for training, and 161 for validation. We also resized each images of the dataset to 256$\times$256. The challenge of this task is that cell type, magnification, and imaging modality have a variety of conditions, so the model need to handle nuclei segmentation in different environment. The second dataset (referred to as GLAND) is from the MICCAI 2015 Gland Segmentation Challenge Contest\cite{sirinukunwattana2015stochastic} (Warwick-QU dataset). In this dataset, there are 165 microscopy gland images in total, 85 images for training, and 60 (Part A) + 20 (Part B) for validation. These training and testing images were collected from healthy glandular tissue and glandular tissue of different degrees of malignancy. We resized each images of the dataset to 512$\times$512, so the training set and testing set are increasing to 158 and 144 respectively.

\begin{figure*}[h]
  \centering
  \includegraphics[width=1.0\linewidth]{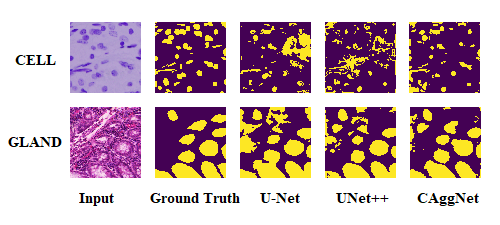} 
  \caption{Visualization of segmentation results of two public datasets.}
  \label{fig:qualitative}
\end{figure*}

We employed pixel-level metrics to evaluate and compare different models for semantic segmentation tasks. These metrics included the mean intersection-over-union (mIoU) score (or Jaccard index similarity) and the harmonic mean of precision and recall (F1-score). 
\begin{equation}
\begin{aligned}
    Pr=\frac{\text{True Positives}}{\text{True Positives}+\text{False Positives}}
\end{aligned}
\end{equation}
\begin{equation}
\begin{aligned}
    Se=\frac{\text{True Positives}}{\text{True Positives}+\text{False Negatives}}
\end{aligned}
\end{equation}
\begin{equation}
\begin{aligned}
    IOU=\frac{Pr*Se}{Pr+Se-Pr*Se}
    \label{eq:IOU}
\end{aligned}
\end{equation}
\begin{equation}
\begin{aligned}
    F_1=2\cdot \frac{Pr\cdot Se}{Pr+Se}
    \label{eq:F1}
\end{aligned}
\end{equation}

where $Pr$ stands for precision and $Se$ stands for sensitivity. 

\subsection{Baselines}
In order to prove the robustness of our model, we compare CAggNet with other semantic segmentation deep learning models, including U-Net\cite{ronneberger2015u}, UNet++\cite{zhou2018unet++} ,and a fully convolutional network (FCN)\cite{long2015fully}.We chose U-Net and UNet++ because their performances are good in many International Conference on Medical Image Computing and Computer Assisted Intervention (MICCAI) contests. The DLA is not in the baseline list because its performance is almost similar to UNet++. In this baseline comparison experiment, all models are trained under the binary cross entropy (BCE) loss.

\begin{equation}
\begin{aligned}
    BCE(x)=-[y\log{f(x)}+(1-y)\log(1-f(X))]
\end{aligned}
\end{equation}

,where $x$ is the input of the model, $f(x)$ is the output probability map and $y\in\{0,1\}$ is the ground truth label.
In table \ref{table:IOU} and table \ref{table:F1}, we compare CAggNet with its baseline models in terms of the IOU and F1-score after repeating the experiment many times. As shown, CAggNet outperforms U-Net, UNet++ and the FCN consistently on both datasets. As a qualitative result, Fig.\ref{fig:qualitative} shows example predictions on the CELL and GLAND validation sets.

\begin{table}[h!]
    \begin{center}
    \begin{tabular}{ |ccc| } 
     \hline
     \multirow{ 2}{*}{Method} & \multicolumn{2}{c|}{Dataset}  \\
     \cline{2-3}
     &CELL&GLAND\\
     \hline
     FCN\cite{long2015fully} & 0.3698 & 0.5400 \\ 
     U-Net\cite{ronneberger2015u} & 0.8459 & 0.7873  \\ 
     UNet++\cite{zhou2018unet++} & 0.8489 & 0.7919\\ 
     CAggNet & \textbf{0.8537} & \textbf{0.7922} \\ 
     \hline
    \end{tabular}
    \caption{Performance of CAggNet and baselines in IOU}
    \label{table:IOU}
    \end{center}
\end{table}

\begin{table}[h!]
    \begin{center}
    \begin{tabular}{ |ccc| } 
     \hline
     \multirow{ 2}{*}{Method} & \multicolumn{2}{c|}{Dataset}  \\
     \cline{2-3}
     &CELL&GLAND\\
     \hline
     FCN\cite{long2015fully} & 0.5293 & 0.6923 \\ 
     U-Net\cite{ronneberger2015u} & 0.9166 & 0.8810  \\ 
     UNet++\cite{zhou2018unet++} & 0.9188 & 0.8836\\ 
     CAggNet & \textbf{0.9216} & \textbf{0.8845} \\ 
     \hline
    \end{tabular}
    \caption{Performance of CAggNet and baselines in F1-score}
    \label{table:F1}
    \end{center}
\end{table}

\subsection{Study on Loss Functions}
To evaluate the contribution of the focal loss, we trained CAggNet with different loss function settings on the GLAND dataset. In table \ref{table:ablation}, we compared the BCE loss function and focal loss function with varying hyperparameters. Models trained with the focal loss generally gain performance against the model trained with BCE loss. By applying a modulating term to the cross-entropy loss, the focal loss allows the model to focus on learning hard positive examples instead of being distracted by the easy negatives, which boosts the performance of our model significantly. Finally, the model trained with focal loss ($\alpha=0.25,\gamma=2$) outperforms the other models and achieves a performance of 0.8063/0.8927 (IoU/F1-score).

\begin{table}[h!]
    \begin{center}
    \begin{tabular}{ |c|c|c|c| } 
     \hline
     Method&IOU&F1-score\\
     \hline
     BCELoss & 0.7922 & 0.8845 \\
     FocalLoss($\alpha=0.5,\gamma=1$) & 0.8020 & 0.8907  \\ 
     FocalLoss($\alpha=0.25,\gamma=2$) & \textbf{0.8063} & \textbf{0.8927}\\ 
     FocalLoss($\alpha=0.5,\gamma=2$) & 0.8048 & 0.8918\\ 
     \hline
    \end{tabular}
    \caption{Study on loss functions on the GLAND dataset}
    \label{table:ablation}
    \end{center}
\end{table}

\subsection{Final Results and Ablation Study}
Table \ref{table:final_results_cell} and table \ref{table:final_results_gland} demonstrates how much the CAM/WAM modules and the focal loss function contributed to the final result. As seen, CAggNet consistently outperforms UNet++ and general U-Net in both the CELL and GLAND datasets. Comparing to U-Net, our proposed architecture achieved an average of 0.94\% gain on F1-score and 1.56\% gain on IOU accuracy. As with UNet++, CAggNet yields average improvement of 0.70\% and 1.18\% points in F1-score and IOU. These results prove the effectiveness of our architecture on both datasets. Fig.\ref{fig:qualitative} shows a qualitative comparison among the results of U-Net, UNet++, and CAggNet.

\begin{table}[h!]
    \begin{center}
    \begin{tabular}{ |c|c|c|c|c|c| } 
     \hline
     U-Net&UNet++&CAggNet&Focal&IoU&F1-score\\
     \hline
     $\surd$ & & & & 0.8459&0.9166 \\
     & $\surd$ & & & 0.8489&0.9188 \\
     & &$\surd$ & &0.8537&0.9216 \\
     & &$\surd$ & $\surd$ &\textbf{0.8581}&\textbf{0.9236} \\
     \hline
    \end{tabular}
    \caption{Final results on CELL dataset}
    \label{table:final_results_cell}
    \end{center}
\end{table}
\begin{table}[h!]
    \begin{center}
    \begin{tabular}{ |c|c|c|c|c|c| } 
     \hline
     U-Net&UNet++&CAggNet&Focal&IOU&F1-score\\
     \hline
     $\surd$ & & & & 0.7873&0.8810 \\
     & $\surd$ & & & 0.7919&0.8836 \\
     & &$\surd$ & & 0.7922&0.8845 \\
     & &$\surd$ & $\surd$ &\textbf{0.8063}&\textbf{0.8927} \\
     \hline
    \end{tabular}
    \caption{Final results on GLAND dataset}
    \label{table:final_results_gland}
    \end{center}
\end{table}

\section{Conclusion}
In this paper, we proposed Crossing Aggregation Network (CAggNet) for medical image segmentation. The suggested architecture consists of two aggregation sub-module: crossing aggregation module (CAM) and weighted aggregation module (WAM). CAM fuses three layers in different levels and computes their residuals as output. And then, groups of CAM connect iteratively in the skip-connection path of U-Net. WAM aggregates the output of the last layer of skip-connections in different levels through the mechanism of channel attention and up-sampling. Experimental results show that the proposed method outperforms several state-of-the-art semantic segmentation methods on two public datasets. The application of focal loss also successfully makes our proposed method avoid the foreground-background imbalance issue in the training process, and finally achieve the highest IoU and F1 score. The plan in the future is to test this model in more medical image segmentation tasks like video frames and CT images. 

\section{ACKNOWLEDGEMENTS}
\label{sec:ACK}

We gratefully acknowledge the support of School of Data Science and School of Basic Medical Sciences at Fudan University for providing NVIDIA 1080Ti GPUs used for this research.






%





\bibliographystyle{plain} 
\bibliography{refs} 

\begin{thebibliography}{10}

\bibitem{fan2020inf}
Deng-Ping Fan, Tao Zhou, Ge-Peng Ji, Yi~Zhou, Geng Chen, Huazhu Fu, Jianbing
  Shen, and Ling Shao.
\newblock Inf-net: Automatic covid-19 lung infection segmentation from ct
  images.
\newblock {\em IEEE Transactions on Medical Imaging}, 2020.

\bibitem{garcia2017review}
Alberto Garcia-Garcia, Sergio Orts-Escolano, Sergiu Oprea, Victor
  Villena-Martinez, and Jose Garcia-Rodriguez.
\newblock A review on deep learning techniques applied to semantic
  segmentation.
\newblock {\em arXiv preprint arXiv:1704.06857}, 2017.

\bibitem{guan2019fully}
Steven Guan, Amir Khan, Siddhartha Sikdar, and Parag Chitnis.
\newblock Fully dense unet for 2d sparse photoacoustic tomography artifact
  removal.
\newblock {\em IEEE journal of biomedical and health informatics}, 2019.

\bibitem{he2016deep}
Kaiming He, Xiangyu Zhang, Shaoqing Ren, and Jian Sun.
\newblock Deep residual learning for image recognition.
\newblock In {\em Proceedings of the IEEE conference on computer vision and
  pattern recognition}, pages 770--778, 2016.

\bibitem{huang2017densely}
Gao Huang, Zhuang Liu, Laurens Van Der~Maaten, and Kilian~Q Weinberger.
\newblock Densely connected convolutional networks.
\newblock In {\em Proceedings of the IEEE conference on computer vision and
  pattern recognition}, pages 4700--4708, 2017.

\bibitem{huang2020unet}
Huimin Huang, Lanfen Lin, Ruofeng Tong, Hongjie Hu, Qiaowei Zhang, Yutaro
  Iwamoto, Xianhua Han, Yen-Wei Chen, and Jian Wu.
\newblock Unet 3+: A full-scale connected unet for medical image segmentation.
\newblock In {\em ICASSP 2020-2020 IEEE International Conference on Acoustics,
  Speech and Signal Processing (ICASSP)}, pages 1055--1059. IEEE, 2020.

\bibitem{iglovikov1801u}
V~Iglovikov and Shvets~A Ternausnet.
\newblock U-net with vgg11 encoder pre-trained on imagenet for image
  segmentation. 2018.
\newblock {\em arXiv preprint arXiv:1801.05746}.

\bibitem{kang2019nuclei}
Qingbo Kang, Qicheng Lao, and Thomas Fevens.
\newblock Nuclei segmentation in histopathological images using two-stage
  learning.
\newblock In {\em International Conference on Medical Image Computing and
  Computer-Assisted Intervention}, pages 703--711. Springer, 2019.

\bibitem{kingma2014adam}
Diederik~P Kingma and Jimmy Ba.
\newblock Adam: A method for stochastic optimization.
\newblock {\em arXiv preprint arXiv:1412.6980}, 2014.

\bibitem{lin2017focal}
Tsung-Yi Lin, Priya Goyal, Ross Girshick, Kaiming He, and Piotr Doll{\'a}r.
\newblock Focal loss for dense object detection.
\newblock In {\em Proceedings of the IEEE international conference on computer
  vision}, pages 2980--2988, 2017.

\bibitem{long2015fully}
Jonathan Long, Evan Shelhamer, and Trevor Darrell.
\newblock Fully convolutional networks for semantic segmentation.
\newblock In {\em Proceedings of the IEEE conference on computer vision and
  pattern recognition}, pages 3431--3440, 2015.

\bibitem{oktay2018attention}
Ozan Oktay, Jo~Schlemper, Loic~Le Folgoc, Matthew Lee, Mattias Heinrich,
  Kazunari Misawa, Kensaku Mori, Steven McDonagh, Nils~Y Hammerla, Bernhard
  Kainz, et~al.
\newblock Attention u-net: Learning where to look for the pancreas.
\newblock {\em arXiv preprint arXiv:1804.03999}, 2018.

\bibitem{ren2015faster}
Shaoqing Ren, Kaiming He, Ross Girshick, and Jian Sun.
\newblock Faster r-cnn: Towards real-time object detection with region proposal
  networks.
\newblock In {\em Advances in neural information processing systems}, pages
  91--99, 2015.

\bibitem{ronneberger2015u}
Olaf Ronneberger, Philipp Fischer, and Thomas Brox.
\newblock U-net: Convolutional networks for biomedical image segmentation.
\newblock In {\em International Conference on Medical image computing and
  computer-assisted intervention}, pages 234--241. Springer, 2015.

\bibitem{shan2020lung}
Fei Shan, Yaozong Gao, Jun Wang, Weiya Shi, Nannan Shi, Miaofei Han, Zhong Xue,
  and Yuxin Shi.
\newblock Lung infection quantification of covid-19 in ct images with deep
  learning.
\newblock {\em arXiv preprint arXiv:2003.04655}, 2020.

\bibitem{sirinukunwattana2015stochastic}
Korsuk Sirinukunwattana, David~RJ Snead, and Nasir~M Rajpoot.
\newblock A stochastic polygons model for glandular structures in colon
  histology images.
\newblock {\em IEEE transactions on medical imaging}, 34(11):2366--2378, 2015.

\bibitem{xia2017w}
Xide Xia and Brian Kulis.
\newblock W-net: A deep model for fully unsupervised image segmentation.
\newblock {\em arXiv preprint arXiv:1711.08506}, 2017.

\bibitem{xiao2018weighted}
Xiao Xiao, Shen Lian, Zhiming Luo, and Shaozi Li.
\newblock Weighted res-unet for high-quality retina vessel segmentation.
\newblock In {\em 2018 9th International Conference on Information Technology
  in Medicine and Education (ITME)}, pages 327--331. IEEE, 2018.

\bibitem{yu2018deep}
Fisher Yu, Dequan Wang, Evan Shelhamer, and Trevor Darrell.
\newblock Deep layer aggregation.
\newblock In {\em Proceedings of the IEEE conference on computer vision and
  pattern recognition}, pages 2403--2412, 2018.

\bibitem{zhang2018exfuse}
Zhenli Zhang, Xiangyu Zhang, Chao Peng, Xiangyang Xue, and Jian Sun.
\newblock Exfuse: Enhancing feature fusion for semantic segmentation.
\newblock In {\em Proceedings of the European Conference on Computer Vision
  (ECCV)}, pages 269--284, 2018.

\bibitem{zhang2019net}
Zhijie Zhang, Huazhu Fu, Hang Dai, Jianbing Shen, Yanwei Pang, and Ling Shao.
\newblock Et-net: A generic edge-attention guidance network for medical image
  segmentation.
\newblock In {\em International Conference on Medical Image Computing and
  Computer-Assisted Intervention}, pages 442--450. Springer, 2019.

\bibitem{zhou2018unet++}
Zongwei Zhou, Md~Mahfuzur~Rahman Siddiquee, Nima Tajbakhsh, and Jianming Liang.
\newblock Unet++: A nested u-net architecture for medical image segmentation.
\newblock In {\em Deep Learning in Medical Image Analysis and Multimodal
  Learning for Clinical Decision Support}, pages 3--11. Springer, 2018.

\end{thebibliography}

\end{document}